\documentstyle[prl,aps,psfig]{revtex}

\draft
\def\cm{\,{\rm cm}}
\def\km{\,{\rm km}}

\def\mpc{\,{\rm Mpc}}

\def\ev{\,{\rm eV}}

\def\gev{\,{\rm GeV}}

\def\sr{\,{\rm sr}}
\def\sec{\,{\rm sec}}

\def\la{\mathrel{\mathpalette\fun <}}
\def\ga{\mathrel{\mathpalette\fun >}}
\def\fun#1#2{\lower3.6pt\vbox{\baselineskip0pt\lineskip.9pt
  \ialign{$\mathsurround=0pt#1\hfil##\hfil$\crcr#2\crcr\sim\crcr}}}
\newcommand{\beq}{\begin{equation}}
\newcommand{\eeq}{\end{equation}}
\begin{document}

\twocolumn[\hsize\textwidth\columnwidth\hsize\csname
@twocolumnfalse\endcsname


\draft
\vspace{-1.5truecm}{\flushright{\hfill hep-ph/9803223}} 
\title{Cosmic Topological Defects, Highest Energy Cosmic Rays, and
the Baryon Asymmetry of the Universe}
   
\author{Pijushpani Bhattacharjee}

\vspace{0.8cm}
\address{Laboratory for High Energy Astrophysics, Code 661,
NASA/Goddard Space Flight Center, Greenbelt, MD 20771, USA.\\}

\vspace{0.8cm}

\address{and}

\vspace{0.8cm}

\address{Indian Institute of Astrophysics, Bangalore - 560 034, INDIA.}

\maketitle

\begin{abstract}
It is pointed out that the observed extremely high energy cosmic rays 
(EHECR) above $10^{11}\gev$ and the observed baryon asymmetry of the
Universe (BAU), both may have a common origin in baryon number
violating decays of supermassive ``X'' particles released from 
cosmic topological defects (TDs) such as cosmic strings and monopoles. 
The X particles produced by TDs in the
recent epochs produce the EHECR, while the BAU is created by X particles 
released from TDs mainly in the very early Universe. 
In this scenario the EHECR is predicted to
contain baryons as well as antibaryons with a small asymmetry between
the two. 
\end{abstract}

\pacs{PACS numbers: 98.80.Cq, 98.70.Sa}
\vskip 2pc]

\noindent 
Cosmic Topological Defects (TDs)\cite{kibble,tdreview}
--- magnetic monopoles, cosmic strings, domain walls, textures,
superconducting cosmic strings, etc.,  
as well as various hybrid systems consisting of these 
TDs --- are predicted to form in the early Universe as a result of
symmetry-breaking phase transitions envisaged in
Grand Unified Theories (GUTs). The TDs can be thought of as `constituted'
of quanta of supermassive gauge and higgs 
fields (generically, ``X'' particles) of the underlying
spontaneously
broken gauge theory, with typical mass $m_X\la10^{16}\gev$, the GUT
symmetry breaking scale. TDs are topologically stable and so,
once formed in the early Universe, they can survive forever with X
particles `trapped' inside them. However, from time to
time, some TDs, through collapse, annihilation or other processes, can
release the trapped X
particles~\cite{bkt,nussinov,hill,hsw,cusp,pbkofu-br-gk,bs,necklace,vincent}. 
Decays of these X particles can give rise to extremely energetic nucleons,
neutrinos and photons~\cite{bhs,abs}
with energies up to $\sim m_X$, which may potentially 
explain~\cite{ssb,slsb,slc-ps-slsc,susytd,bbv} the  
Extremely High Energy Cosmic Ray (EHECR) events   
with energies above $10^{11}\gev$~\cite{ehecrevents}. 

The energies associated with the EHECR events are hard to
obtain~\cite{ssb,hillas-norman} within conventional scenarios
of acceleration of charged particles in relativistic shocks
associated with powerful astrophysical objects.  
In addition, there is the problem of
absence of any obviously identifiable sources for the EHECR
events~\cite{ssb,es}. 
These problems are avoided in the TD scenario in a natural way. Firstly,
no acceleration mechanism is needed: The decay products of the X particles
have energies up to $\sim m_X$ which can be as large as 
$\sim 10^{16}\gev$. 
Secondly, the absence of obviously identifiable sources is not a
problem because TDs need not necessarily be associated with any
visible or otherwise active astrophysical objects. 

The purpose of this Letter is to point out that if indeed decays of X
particles from one or more TD-processes are 
responsible for the observed EHECR, then the same TD-processes may also be
responsible for creating the Baryon Asymmetry of the Universe (BAU)
due to CP- and baryon number ($B$) violating decays of the X particles.  
Production of X particles from TDs is an irreversible process, so the
out-of-thermal-equilibrium condition necessary for the creation of BAU is
(and hence the Sakharov conditions are) automatically satisfied. 

The total baryon asymmetry produced by the decays of all X
particles released from TDs at all epochs in the past is calculated below  
by normalizing the X particle production rate in the
present epoch to that required to explain the EHECR flux.   
The result of this exercise is that the observed BAU
can be obtained provided the temperature $T_F$,
defined as the temperature below
which any $B$-asymmetry produced by the X particle decays is not erased by
other $B$-violating processes, is $\sim 10^{14}\gev$. 

Of course, as is well-known, $B$ violation
through sphaleron transitions at high temperatures~\cite{sphaleron} could
erase any net $B$-asymmetry created by other processes {\it unless} a
non-vanishing value of $B-L$ ($L$ being the Lepton number) is generated.
This is a general problem for any scenario of so-called ``GUT
baryogenesis'' in which BAU is generated through decay of supermassive
(GUT-scale) X particles at high temperatures above the electroweak scale
(few hundred GeV). The way to avoid this problem is also well-known:
Assume a GUT like $SO(10)$, in which a net $B-L$ asymmetry may be
generated through $L$-violating decays of certain Higgs
particles (see, e.g., Refs.~\cite{kt,baureview} for reviews of various
baryogenesis scenarios); we shall assume this to be the case.       

The possibility that $B$ violating decays of X particles
released from TDs could be responsible for the BAU was first pointed out 
in two independent works~\cite{bkt,nussinov} in 1982, 
and subsequently studied further in Ref.~\cite{bdh}. However, the
possibility that TDs might be relevant for EHECR was not explored then.
On the other hand, although there is currently much interest
in the possibility that TDs may be responsible for EHECR, the
possibility that the same TD processes might also have been responsible
for the BAU seems to have escaped notice thus far. 
Thus, the present Letter explores the possible link
between EHECR and BAU and a possible common origin of both in
decays of X particles released from TDs. 

The number density of X particles produced by TDs per unit time,
$dn_X/dt$, can be generally written as\cite{bhs}
\beq
{dn_X\over dt}(t)= {Q_0\over m_X} \left({t\over
t_0}\right)^{-4+p}\,,\label{xrate}
\eeq
where $t_0$ denotes the present epoch, and $Q_0$ is the rate of energy
density injected in the form of X particles in the present epoch. The
quantity $Q_0$ and the
parameter $p$ depend on the specific TD process under consideration. 
Processes with $p < 1$ generally lead to unacceptably
high rate of energy injection in the early cosmological epochs 
which would cause excessive ${}^4$He photo-disintegration and CMBR
distortions~\cite{sigljedam} and are, therefore, unfavored in the context
of EHECR. We shall, therefore, consider only the case $p=1$, which is 
representative of a number of TD processes  
studied so far~\cite{pbkofu-br-gk,bs,necklace,vincent} that are
likely to be relevant for EHECR.  

Although Eq.~(\ref{xrate}) is valid 
at all epochs after the formation of the relevant TDs in the
early Universe, only those X
particles produced in the relatively recent epochs and at relatively
close-by, non-cosmological distances ($\la100\mpc$) are relevant for the
question of EHECR. This is because protons above
$10^{11}\gev$ produce pions in collision with the photons of the 
cosmic microwave background radiation (CMBR), and as a result 
suffer drastic energy loss (the so-called ``GZK
effect''~\cite{gzk}), which limits the source distance to effectively only
a few tens of Mpc~\cite{ssb,es,ahar-cronin}. 
Distances of sources of photons of energies above
$\sim10^{11}\gev$ are also similarly restricted  
due to absorption through $e^+e^-$ production on the radio background
photons (see, e.g., Refs.~\cite{abs,bbv}). The neutrinos can
survive from much earlier cosmological epochs; however, the detected EHECR
events are unlikely to be due to neutrinos because of their much lower
interaction cross-section compared to those of nucleons and photons.  
Thus the EHECR can be produced, if at all, by X particles
released from TDs only in the very recent epochs. 

The BAU, on the other hand, must already have been in place in the
early Universe; indeed, in order not to disturb the successful
predictions of the primordial nucleosynthesis scenario, the bulk of
the BAU must have been created before 
a temperature of $\sim$ 0.1 MeV, and probably well before the quark-hadron
phase transition at a temperature of few hundred MeV. 
This is naturally achieved in the TD scenario because, the X particle
production rate being $\propto t^{-3}$, the bulk of the contribution to 
the BAU comes from 
production and decay of X particles at the earliest possible epoch
characterized by the temperature $T_F$ mentioned earlier. 

The X particles (including $\bar{X}$'s) released from TDs decay 
typically into quarks and leptons (and/or their antiparticles). 
The quarks ``fragment'' into jets of hadrons --- mostly 
pions, with a small admixture (typically
$\la10\%$) of baryons and antibaryons (nucleons and antinucleons). 
Thus photons and neutrinos resulting from the decays of neutral and
charged pions, respectively, dominate the total particle yield at
production~\cite{fn1}, while  
the baryon asymmetry associated with the X-decay is contained in the
relatively small baryon yield.  

The hadronic spectra should be similar to those measured for jets seen 
in $e^+e^-$ annihilation experiments, which are well described by
QCD~\cite{fragmentation}. For the energy regions of our interest, the 
nucleon, photon and neutrino spectra resulting from the decay of each X 
particle can be approximated by power-laws in energy
($\propto E^{-\alpha}$) with, typically, $1.3\la\alpha\la1.7$. 

We can now make a rough estimate of $Q_0^{\rm EHECR}$, the value of
$Q_0$ in Eq.~(\ref{xrate}) required to explain the observed EHECR flux. 
We can do this by normalizing the
predicted photon flux with the observed EHECR flux 
since, in our scenario, photons dominate the observable
particle flux at EHECR energies. 
Let us assume a typical $B$-violating decay mode of the X into a
quark and a lepton: $X \to q\ell$. The quark produces a hadronic jet. 
The photons from the decay of neutral pions in the jet carry a total
energy
$E_{\gamma,{\rm Total}}\simeq
\left({1\over3}\times0.9\times{1\over2}\right)m_X = 0.15 m_X$, where we
have assumed that on average $\sim$ 90\% of a jet's total energy is
carried by pions. Assuming a power-law photon
spectrum with index 1.5, the photon injection spectrum due to a single X
particle decay can be written as 
${dN_\gamma\over dE_\gamma} = {1\over m_X} \times 0.3 \, \left({2
E_\gamma\over m_X}\right)^{-1.5}$, 
which is properly normalized with the total photon energy $E_{\gamma,{\rm
Total}}$. We can neglect cosmological evolution effects and take the 
present epoch values of the relevant quantities, since photons of
EHECR energies have a cosmologically negligible absorption length of only
$\sim$ few tens of Mpc. 
With these assumptions, the photon flux $j_\gamma (E_\gamma)$ 
is simply given by 
$j_\gamma (E_\gamma)\simeq {1\over 4\pi} \lambda(E_\gamma)\, {dn_X\over
dt}\, {dN_\gamma\over dE_\gamma}$, 
where $\lambda(E_\gamma)$ is the pair production absorption path length of
a photon of energy $E_\gamma$. 
 
Noting that $dn_X/dt = Q_0/m_X$, and normalizing the above flux to the
measured EHECR flux corresponding to the highest energy event at $\sim
3\times10^{11}\gev$, given by $j(3\times10^{11}\gev)\approx
5.6\times10^{-41}\cm^{-2} \ev^{-1} \sec^{-1} \sr^{-1}$~\cite{ehecrevents},
we get 
\beq
Q_0^{\rm EHECR}\approx 1.2\times10^{-30}{\gev\over\cm^3\sec}
\left({30\mpc\over\lambda_{\gamma,300}}\right)
\left({m_X\over10^{16}\gev}\right)^{1\over2}\,,\label{Q_0_required}
\eeq
where $\lambda_{\gamma,300}$ is the absorption path length of a 300 EeV
photon (1 EeV $\equiv 10^{18}\ev$). The subscript 0 stands for the present
epoch. 
More detailed numerical calculations of the predicted EHECR flux   
have been done~\cite{slc-ps-slsc} by solving the relevant full transport
equations, which yield a
similar number for $Q_0^{\rm EHECR}$ as derived above. 

Eq.~(\ref{Q_0_required}) implies a required X particle
production rate $\dot{n}_{X,0}\sim 1.1 
\times 10^{35} \left({30\mpc\over\lambda_{\gamma,300}}\right) 
\left({m_X\over10^{16}\gev}\right)^{-1/2} \mpc^{-3} {\rm yr}^{-1}$. 
In addition, in order that the resulting EHECR flux be not too
anisotropic, there must
be several TD sources of these X particles today within a typical ``GZK''
volume of radius $\sim 30\mpc$. This puts rather stringent constraints on
the possible types of TD sources~\cite{bbv}. 
Currently, the most viable TDs in this regard seem to be 
collapsing monopolonia~\cite{hill,bs} and ``necklace''~\cite{necklace},
although X particle production due to repeatedly
self-intersecting kinky cosmic string loops~\cite{siemens-kibble} also
remains as a
possibility. Empirically, in terms of the number densities of TDs, 
perhaps the most well-constrained are the monopoles~\cite{kt}.
Noting that a GUT monopole typically has a mass
$m_M\sim 40m_X$, so that each monopolonium collapse releases $\sim 80$ X
particles, the above estimate of required $\dot{n}_{X,0}$ implies the
condition $(n_{M\bar M}^c/n_M)\Omega_M h^2\simeq 1.2\times10^{-8}
\left({30\mpc\over\lambda_{\gamma,300}}\right)
\left({m_X\over10^{16}\gev}\right)^{1/2} (\Omega_0 h^2)^{-1/2}$, where 
$n_{M\bar M}^c$ is the number density of monopolonia in the
final stage of collapse in the present epoch, i.e., the ones
that are currently, and will be (over the next one Hubble time), 
disappearing due to $M\bar M$ annihilation~\cite{fn3}, $n_M$ is the number
density
of monopoles (including antimonopoles), $\Omega_M$ is the mass density
contributed by monopoles and $\Omega_0$ the total mass density (both in
units of the critical mass density of closure of the Universe), and  
$h$ is the present Hubble constant in units of $100\km \sec^{-1}
\mpc^{-1}$. An equilibrium Saha ionization formalism~\cite{hill,bs} for 
the monopolonium formation process indicates that the
above requirement on $(n_{M\bar M}^c/n_M)$ is consistent with the
known independent upper limits on the monopole abundance such as the
closure limit $\Omega_M h^2\leq 1$, and also with the more stringent
``Parkar limit''~\cite{kt} $(\Omega_M h^2)_{\rm
Parkar}\la4\times10^{-3}(m_M/10^{16}\gev)^2$. 
For a recent discussion of phenomenological aspects of various TD
sources of EHECR, see Ref.~\cite{bbv}. 

Now, taking $Q_0=Q_0^{\rm EHECR}$ in
Eq.~(\ref{xrate}) with $p=1$, the total BAU produced by X particle 
decays is simply given by~\cite{bkt} 
\beq
{n_B\over s}\simeq\Delta{B}\int\limits_{t_F}^{t_0} {dt\over s}{dn_X\over
dt}(t)= \Delta{B} {Q_0^{\rm EHECR}\over m_X} \int\limits_{t_F}^{t_0}
{dt\over s}\left({t_0\over t}\right)^{3}\,,\label{B1}
\eeq
where $t_F$ is the ``freeze-out'' time corresponding to the temperature
$T_F$ mentioned earlier, $\Delta{B}$ is the mean net baryon number
produced in the decay of an X particle, and 
$s=(2\pi^2/45) g_{*S} T^3$ is the entropy density at
temperature $T$ (corresponding to time $t$), $g_{*S}\sim100$ being the
relevant number of relativistic degrees of freedom~\cite{kt}. In writing
Eq.~(\ref{B1}), we have assumed that the entropy produced in the decays of
X particles themselves at any time $t$ in the early epochs of
our interest is negligible compared to the ambient entropy density.
This is a good approximation~\cite{bkt} in our scenario because
the X particles do not dominate the density of the Universe at any time. 

Using the standard time-temperature relation in the early
Universe~\cite{kt}, we see that the integral in
Eq.~(\ref{B1}) is dominated by the contribution at the earliest time
$t_F$, giving 
\begin{eqnarray}
{n_B\over s}\simeq & 2.2\times10^{-9}\left(\Delta{B}/
10^{-2}\right) 
\left(m_X/10^{16}\gev\right)^{-1/2} \nonumber \\
& \, \left(T_F/10^{16}\gev\right) \Omega_0^{-3/2}
(h/0.65)^{-3}\,,\label{B2}
\end{eqnarray}
where 
 we have
used $\lambda_{\gamma,300}= 30\mpc$ in Eq.~(\ref{Q_0_required}). 

In a general GUT-baryogenesis scenario, the temperature $T_F$ is $\sim
m_X$~\cite{kt,baureview}. We thus see that, depending on the values of
$\Delta{B}$, $h$ and $\Omega_0$, the estimated BAU of
$\sim$ (4 -- 14) $\times 10^{-11}$~\cite{kt} can
be obtained with $m_X\sim10^{14}\gev$. (Note that $m_X$ much above
$10^{14}\gev$ tends to overproduce BAU, as well as the
EHECR~\cite{susytd}, and may, therefore, be unfavored.)  

There are many uncertainties in the above estimates. An accurate estimate
can be obtained only through detailed numerical calculations involving
solution of the relevant Boltzmann equation that incorporates all
B-violating interactions, not just the decays of X's. Nevertheless, the
rough
estimate of BAU made above should give us an idea of the kind of numbers
to expect. 

Recently, attempts have been made~\cite{preheating-X} to revive the
``standard'' GUT baryogenesis through decay of massive X particles
produced during ``preheating'' stage 
in the inflationary Universe. While these X particles may produce the  
BAU, they cannot be responsible for the EHECR because these X particles
all decayed away in the early Universe and are
not produced in the recent epochs. On the other hand, it has been
suggested~\cite{stable-X}
that massive {\it stable} particles (with life-time $\ga$ age of the
Universe) may be created in the early Universe through vacuum fluctuations  
during inflation, which can act as the dark matter, and a small fraction
of those decaying in the present epoch may give rise to the EHECR.
However, in this scenario, the BAU (which must have been created in the
early Universe) cannot have an origin in decays of these stable X
particles, which decay, if at all, only in the recent epochs. In contrast,
in the TD scenario outlined above, the X
particles themselves are unstable, but they are {\it produced continually 
at all epochs} including the recent epochs, so that {\it both} BAU as well
as EHECR can be produced. 

Note that, in the TD scenario, the EHECR at production is predicted to
contain baryons as well as antibaryons with a small asymmetry between the
two. It remains, however, as a challenge at this stage to devise a scheme
that would enable one to distinguish EHECR air-showers initiated by
protons from those initiated by antiprotons, thereby  
to test the prediction experimentally. 

To summarize, then, not only the extremely {\it high energy} cosmic rays,
but the entire `low' energy baryonic content of the Universe may,
at some stage or other, have originated from topological defects. 
Thus the BAU may be a dynamically evolving quantity, and 
the EHECR observed today may represent the baryon creation process
itself ``in action'' in the Universe today. 

I wish to thank Q.~Shafi, F.~Stecker, R.~Streitmatter, and G.~Yodh for
discussions. This work is supported by NAS/NRC and NASA.


\begin{references}
\bibitem{fn1} However, protons may still dominate over photons in the
observed particle flux at energies $\la10^{11}\gev$ because of the
larger attenuation length of protons compared to the absorption length
of photons at these energies, while the photons can dominate over
protons above a few times $10^{11}\gev$ (see Refs.~\cite{abs,bbv}).
\bibitem{fn3} These monopolonia were formed over a Hubble time roughly at
about the epoch of primordial nucleosynthesis~\cite{bs}.
\end{references}
\end{document}